\documentstyle[prb,aps,eqsecnum]{revtex}

\begin{document}

\draft

\title{Renormalization group approach to layered superconductors}
\author{C. Timm}
\address{Universit\"at Hamburg, I. Institut f\"ur Theoretische
Physik, Jungiusstrasse 9, D-20355 Hamburg, Germany}
\date{July 4, 1995}

\maketitle

\hfill supr-con/9507002 \hspace{1.8cm}

\begin{abstract}
A renormalization group theory for a system consisting of coupled
superconducting layers as a model for typical high-temperature
superconducters is developed. In a first step the
electromagnetic interaction over infinitely many layers is taken into
account, but the Josephson coupling is neglected. In this case the
corrections to two-dimensional behavior due to the presence of
the other layers are very small. Next,
renormalization group equations for a layered system
with {\it very strong\/} Josephson coupling are derived, taking into
account only the smallest possible Josephson vortex
loops. The applicability of these two limiting cases
to typical high-temperature superconductors is discussed.
Finally, it is argued that the original
renormalization group approach by Kosterlitz is not applicable
to a layered system with {\it intermediate\/} Josephson coupling.
\end{abstract}


\narrowtext

\section{Introduction}
\label{sec:intro}

Since the highly anisotropic high-temperature superconductors
(HTSC's) consist of weakly coupled superconducting layers one expects
two-dimensional effects to be important.
Quasi-two-dimensional superconducting films near the critical
temperature and in the absence of an external magnetic
field are well described by the
Berezinskii-Kosterlitz-Thouless (BKT) theory.\cite{BKT}
The BKT theory states that the unbinding of spontaneously created
vortex-antivortex pairs is responsible for the phase transition of
these films. A rigorous renormalization group formulation of the BKT
theory was given by Kosterlitz.\cite{Kost}
The question arises of whether the BKT theory can be extended to
describe layered superconductors.

In this paper we first consider a system of superconducting layers
with exclusively electromagnetic coupling and than extend our approach
to include the Josephson coupling between the layers. The main
consequence of the Josephson coupling is the appearance of Josephson
vortices (JV's), i.e.\ vortex strings that reside {\it between\/} the
superconducting layers (for a review see Blatter
{\it et al.}\cite{Blat}).

The layered system without Josephson coupling has been
treated by Scheidl and Hackenbroich\cite{Sche} within a selfconsistent
linear response theory.
The authors describe the renormalization of the interaction between
vortices residing in the same layer, in neighboring layers, etc.,
by a single equation. It is shown below, however, that the screening
cannot be described in such a simple way.

Several authors have considered the Josephson coupled
case.\cite{Horo,Fisch,Pier,Frie,Chat} Horovitz\cite{Horo} has
studied the phase transitions due to a) the unbinding of
vortex-antivortex pairs with vanishing Josephson coupling
and b) JV loops that reside entirely {\it between\/} the layers
within a sine-Gordon renormalization group approach. The
competition between these two mechanisms is then discussed in a
partly heuristic manner. The role of JV loops that cross the
layers does not become clear.

Fischer\cite{Fisch} has
assumed that the renormalization of the
linear term in the interaction of pancake vortices, which is due to
the presence of JV's, and the
logarithmic, electromagnetic term can be described by the same
screening factor. This is not the case.

Pierson\cite{Pier} has used a renormalization group approach
similar to Ref.~\onlinecite{Kost} to treat the Josephson coupled
system. However, the partition
function given by Pierson does not correctly describe a layered
superconductor because it contains a linear term in the vortex
interaction for any two pancake vortices in the same layer or in
neighboring layers, whereas such a term should only be present for
two pancakes that are actually connected by a JV.

Friesen\cite{Frie} has considered
the further simplified model of one superconducting layer between two
thick superconducting slabs, using
an expression for the vortex interaction
which is correct for small separations only. This
approximation is not sufficient since in the renormalization procedure
the interaction between widely separated
vortices must be taken into account.

A different approach to the same problem has been used by
Chattopadhyay and Shenoy.\cite{Chat} They start from the
anisotropic three-dimensional (3D) XY model on a cubic lattice.
There are two reasons, however, why we prefer to use a different model:
1.\ The HTSC's are essentially discrete in one direction, but continuous
in the other directions, whereas the XY model of Ref.\ \onlinecite{Chat}
is discrete in all directions. Universality ensures that both systems
behave similarly within the 3D critical region, but the present
paper is concerned with the 2D behavior {\it outside\/} that region,
where universality does not hold.
2.\ The XY model on a cubic lattice contains three independent parameters
(two spin couplings and a lattice constant), whereas the model considered
here contains five parameters in the case of non-vanishing Josephson
coupling.
Experiments suggest that there are no universal relations between these
parameters, i.e.\ they are truly independent. Thus the 3D XY model
appears to be insufficient to describe the HTSC's.
Furthermore, like Pierson\cite{Pier} and Friesen\cite{Frie} the authors
make the approximation of a linear term in the
interaction between any two pancake vortices, irrespective
of the positions of the JV's.

The present paper is organized as follows: In Sec.~\ref{sec:ea} a
rigorous renormalization group theory for the exclusively
electromagnetically coupled system is developed and
the quantitative predictions of this theory are discussed for a
typical HTSC. This theory then serves as a reference frame for the
further discussion. In Sec.~\ref{sec:Joe}
the consequences of Josephson coupling are investigated.
The paper concludes with a brief summary.

\section{The electromagnetic approximation}
\label{sec:ea}

\subsection{Derivation of the renormalization group equations}
\label{suse:ren}

Let us now turn to the layered system with negligible Josephson
coupling. This approximation may be applicable to
artificially grown HTSC superlattices since these structures
are characterized by a large layer separation compared to the
coherence length perpendicular to the layers.

The unrenormalized interaction between two vortices
separated by a distance $r$ within the same
layer is given by\cite{Buzd}
\begin{equation}
U_0(r) = \mp \frac{\phi_0^2 s}{8\pi^2 \lambda_{ab}^2}
  \ln\frac{r}{\tau} ,
\end{equation}
where the upper (lower) sign is for vortices of the same (opposite)
vorticity, $\phi_0 = hc/2e$ is the flux quantum, $s$ is the layer
separation, $\lambda_{ab}$ is the
magnetic penetration depth for fields applied perpendicular to the
layers, and $\tau$ is the minimum vortex separation.
The interaction between vortices that are $n\not=0$ layers apart
is\cite{Arte}
\begin{equation}
U_n(r) = \pm \frac{\phi_0^2 s^2}{16\pi^2 \lambda_{ab}^3}
  \exp\left(-\frac{|n|\,s}
  {\lambda_{ab}}\right) \ln\frac{r}{\tau} .
\end{equation}
Note the opposite sign and the reduction by the small factor
$s/2\lambda_{ab}$ compared to the interaction within the same layer.
Note further that the interaction of vortices that are, say, 2 or 3
layers apart is of the same order of magnitude as the interaction
for $n=1$. Thus it would not be justified to include only the
interaction between vortices in neighboring layers and neglect
terms with $|n|>1$.

Defining the vortex ``charge''
\begin{equation}
q = \pm \sqrt{\frac{\phi_0^2 s}{8\pi^2 \lambda_{ab}^2}}
\end{equation}
and coupling parameters
\begin{equation}
\alpha_n = \left\{
  \begin{array}{ll}
    -1 & \quad\mbox{for $n=0$} \\
    \displaystyle\frac{s}{2\lambda_{ab}}
    \exp\left(-\frac{|n|\,s}{\lambda_{ab}}\right) & \quad
    \mbox{for $n\not=0$}
  \end{array} \right. ,
\label{def:an}
\end{equation}
we can write down the grand canonical partition function,
\begin{eqnarray}
{\cal Z} & = & \sum_N \frac{1}{N!^2} \left(\frac{z}{\tau^2}\right)^{2N}
  \sum_{n_1} \int_{D_1} \!\!d^2r_1 \cdots \sum_{n_{2N}}
  \int_{D_{2N}} \!\!d^2r_{2N} \nonumber \\
& & \times \exp\!\Bigg(-\frac{\beta}{2} \sum_{i\not=j}
  q_i q_j \alpha_{n_i-n_j} \ln\frac{|{\bf r}_i-{\bf r}_j|}{\tau}
  \Bigg) .
\label{Z0}
\end{eqnarray}
Here, $N$ is the total number of vortex-antivortex pairs in the system,
$z=\exp(\beta\mu)$ is the vortex fugacity, $\beta=1/k_BT$ is the
inverse temperature, $n_i$ and ${\bf r}_i$ are
the number of the layer and the position within the layer of vortex
$i$, respectively, and
the range of integration $D_i$ is the whole
of the layer $n_i$ except discs of radius $\tau$ centered at
the vortices $j<i$. Since the external magnetic field vanishes
the sums over the layer indices $n_i$ are
subject to the constraint that the total vorticity in every layer
be zero.

Following the program of the renormalization group approach,\cite{Kost}
we now proceed to integrate out the smallest vortex-antivortex pairs
(of size between $\tau$ and $\tau+d\tau$).
The central concept is to rewrite the
resulting expression in such a way that it has the same functional form
as the original partition function, but with renormalized
parameters. In our case, these parameters are the fugacity $z$ and
the coupling parameters $\alpha_n$. As more and more pairs are
integrated out, the parameters are renormalized to incorporate the
effect of these pairs on the system. The renormalization is
described by differential equations for the parameters as functions of
the smallest pair size $\tau$. The properties of the macroscopic system
are governed by the limiting behavior of these recursion relations
for large $\tau$.

In the case of a single layer\cite{Kost} only neutral pairs are
integrated out so that the total vorticity remains zero. A change
in the total vorticity would result in an infinite contribution to
the energy and could not be compensated for by suitable
renormalization of parameters.
Similarly, in the present case the vorticity in {\it every\/}
layer must remain zero because removal of any configuration
of vortices that does not consist of pairs in the layers would
make the energy of the remaining system infinite.
(This observation is related to the fact that the interaction
is screened by vortices which move freely within the planes, but
not perpendicular to the planes.)
Thus only pairs which reside in the same layer can be integrated out.
This limitation of the present approach is expected to have unphysical
consequences, especially concerning the correlations between pancakes
in different layers, which are essential within the (narrow) 3D
critical region.

\widetext
The integration procedure
is similar to the case of a single layer.\cite{Kost} It may be
written as the following prescription:
\begin{eqnarray}
\lefteqn{
\sum_{n_1} \int_{D_1} \!\!d^2r_1 \cdots \sum_{n_{2N}} \int_{D_{2N}}
  \!\!d^2r_{2N} \cong \sum_{n_1} \int_{D'_1} \!\!d^2r_1 \cdots
  \sum_{n_{2N}} \int_{D'_{2N}} \!\!d^2r_{2N} } \nonumber \\
& & {}+\frac{1}{2} \sum_{i\not=j} \sum_{n_1} \int_{D'_1} \!\!d^2r_1
  \cdots \sum_{n_{i-1}} \int_{D'_{i-1}} \!\!d^2r_{i-1} \sum_{n_{i+1}}
  \int_{D'_{i+1}} \!\!d^2r_{i+1} \cdots
  \sum_{n_{j-1}} \int_{D'_{j-1}} \!\!d^2r_{j-1} \sum_{n_{j+1}}
  \int_{D'_{j+1}} \!\!d^2r_{j+1} \cdots \nonumber \\
& & \quad\times \sum_{n_{2N}} \int_{D'_{2N}} \!\!d^2r_{2N}
  \sum_{n_j} \int_{\overline{D}_j} \!\!d^2r_j
  \sum_{n_i} \int_{\tau\le |{\bf r}_i-{\bf r}_j|<\tau+d\tau} \!\!\!\!
  d^2r_i \delta_{n_i,n_j} \delta_{q_i,-q_j} .
\label{intout}
\end{eqnarray}
Here, the $D'_i$ are the same as the $D_i$ above, but with $\tau$
replaced by $\tau+d \tau$, and $\overline{D}_j$ is the whole layer
except discs of radius $\tau$ centered at the vortices $k\not=i,j$. The
sum over $n_i$ can be performed trivially.
The right-hand
side of Eq.~(\ref{intout}) consists of two summands: The first one
is similar to the whole left-hand side, but with minimum pair size
$\tau+d\tau$ instead of $\tau$. The second one gives the approximate
correction.
The last integral in Eq.~(\ref{intout}) integrates over the separation
vector of a small pair. The integral over ${\bf r}_j$ and the sum over
$n_j$ take the pair over the whole layer and all the layers,
respectively. The sum $1/2 \sum_{i\not=j}$
selects every possible pair just once.

We now proceed to apply the prescription (\ref{intout}) to the
partition
function (\ref{Z0}). The mathematical procedure is similar to the
one of Ref.\ \onlinecite{Kost}, taking into account the additional
sums over layer indices. The result is
\begin{eqnarray}
{\cal Z} & = & \exp\left[2\pi\left(\frac{z}{\tau^2}\right)^2
  \tau\,d\tau
  MF\right] \sum_N \frac{1}{N!^2} \left(\frac{z}{\tau^2}\right)^{2N}
  \sum_{n_1} \int_{D'_1} \!\!d^2r_1 \cdots \sum_{n_{2N}} \int_{D'_{2N}}
  \!\!d^2r_{2N} \nonumber \\
& & \times \exp\!\Bigg[ -\frac{\beta}{2} \sum_{i\not=j}
  \left(\alpha_{n_i-n_j}
  + 2\pi^2 z^2\frac{d\tau}{\tau} \beta q^2 \tilde{\alpha}_{n_i-n_j}
  \right) q_iq_j \ln\frac{|{\bf r}_i-{\bf r}_j|}{\tau} \Bigg] ,
\end{eqnarray}
where $\tilde{\alpha}_n = \sum_m \alpha_{m-n}\alpha_m$ and $F$ and $M$
are the size and the number of the layers, respectively.
Up to this point, $\tau$ has been replaced by $\tau+d\tau$ only in the
ranges of integration $D'_i$. However, we have to rescale $\tau$
everywhere to be consistent. To order $d\tau$ we eventually obtain
\begin{eqnarray}
{\cal Z} & = & \exp\left[2\pi\left(\frac{z}{\tau^2}\right)^2
  \tau\,d\tau MF\right] \sum_N \frac{1}{N!^2}
  \left(\frac{z}{(\tau+d\tau)^2}\right)^{2N}
  \left[1+\Bigl(2+\frac{\beta}{2}q^2\alpha_0\Bigr)\frac{d\tau}{\tau}
  \right]^{2N} \nonumber \\
& & \times \sum_{n_1} \int_{D'_1} \!\!d^2r_1 \cdots
  \sum_{n_{2N}} \int_{D'_{2N}} \!\!d^2r_{2N}
  \exp\!\Bigg[-\frac{\beta}{2} \sum_{i\not=j} \left(\alpha_{n_i-n_j}
  + 2\pi^2 z^2\frac{d\tau}{\tau} \beta q^2 \tilde{\alpha}_{n_i-n_j}
  \right)
  q_iq_j \ln\frac{|{\bf r}_i-{\bf r}_j|}{\tau+d\tau} \Bigg] .
\label{ZE}
\end{eqnarray}

\narrowtext
Note that no length parallel to the $z$ axis is rescaled,
in contrast to the length scale in the planes, $\tau$. Indeed,
the partition function (\ref{Z0}) does not contain any length scale in
the $z$ direction.
We can interpret the layer index as an internal, discrete degree of
freedom. The system we get in this way is equivalent to the layered model
since both have the same partition function, but it is two-dimensional.
Thus, rescaling of lengths in the $z$~direction is meaningless for both
systems.

If we compare Eq.\ (\ref{ZE}) with the original partition function
(\ref{Z0}), we find that both have indeed the same functional form.
Dropping the irrelevant factor $\exp(2\pi(z/\tau^2)^2 \tau d\tau MF)$,
the partition function takes the form of the original partition
function if we set
\begin{equation}
z \to \left[1+\Bigl(2+\frac{\beta}{2}q^2\alpha_0\Bigr)
  \frac{d\tau}{\tau}\right] z
\end{equation}
and
\begin{equation}
\alpha_n \to \alpha_n + 2\pi^2 z^2 \frac{d\tau}{\tau} \beta q^2
  \tilde{\alpha}_n .
\end{equation}

The same information can be expressed as a system of coupled
differential equations:
\begin{eqnarray}
\frac{dz}{dl} & = & z\left(2+\frac{\beta}{2} q^2 \alpha_0\right) , \\
\frac{d\alpha_n}{dl} & = & 2\pi^2 z^2 \beta q^2 \tilde{\alpha}_n ,
\end{eqnarray}
where a logarithmic length scale $l=\ln \tau/\tau_0$ is introduced.
Here, $\tau_0$ is the unrenormalized minimum pair separation.
These equations may be rewritten as
\begin{eqnarray}
\frac{dz^2}{dl} & = & z^2(4+\beta q^2 \alpha_0) ,
\label{Gz2} \\
\frac{d\alpha_n}{dl} & = & 2\pi^2 z^2 \beta q^2 \sum_m \alpha_{m-n}
  \alpha_m .
\label{Gan}
\end{eqnarray}
This infinite set of equations replaces the Kosterlitz
recursion relations in the case of an electromagnetically coupled
layered system.
It is subject to the boundary condition that $z^2$ and $\alpha_n$
take on their bare values for $l=0$.
Note that by letting
$\alpha_0=-1$ and $\alpha_n=0$ for $n\not=0$
we regain the Kosterlitz recursion relations.

\subsection{Weak electromagnetic coupling}
\label{suse:weak}

The macroscopic properties of the layered superconductor are controlled
by the behavior of the solution of Eqs.~(\ref{Gz2}) and (\ref{Gan}) for
large length scales $l$. These equations can be simplified by means of
an expansion for small electromagnetic interaction between vortices in
different layers.

To this end we define the quantity
\begin{equation}
{\cal A}^2 = \sum_{n\not=0} \frac{\alpha_n^2}{\alpha_0^2} ,
\end{equation}
which is a measure for the coupling between different layers as
compared with the coupling within the same layer.
{}From the definition (\ref{def:an}) it follows that in the
unrenormalized case ${\cal A}^2 \cong s/4\lambda_{ab} \ll 1$
for all HTSC's. We will see below that this inequality holds for the
renormalized quantities also.

If we further define the usual stiffness constant
\begin{equation}
K(l) = - \frac{\alpha_0(l) \beta q^2}{2\pi} ,
\end{equation}
we find, to linear order in ${\cal A}^2$,
\begin{eqnarray}
\frac{dz^2}{dl} & = & 2z^2(2-\pi K) ,
\label{GG1} \\
\frac{dK}{dl} & = & -4\pi^3 z^2 K^2 (1+{\cal A}^2) ,
\label{GG2} \\
\frac{d{\cal A}^2}{dl} & = & -8\pi^3 z^2 K {\cal A}^2 .
\end{eqnarray}
Formal integration of the last equation yields
\begin{equation}
{\cal A}^2(l) = {\cal A}_0^2 \exp\!\bigg[- 8\pi^3 \int_0^l \! dl'\,
  z^2(l') K(l') \bigg] ,
\label{Aint}
\end{equation}
where ${\cal A}_0^2 = {\cal A}^2(l=0)$.

Since ${\cal A}_0^2 \ll 1$ we may expand the
square of the fugacity and the stiffness
constant to linear order in ${\cal A}_0^2$,
\begin{eqnarray}
z^2 & = & z_0^2 + \Delta z^2 \, {\cal A}_0^2 ,
\label{exp1} \\
K & = & K_0 + \Delta K \, {\cal A}_0^2 .
\end{eqnarray}
To the same order, $z^2$ and $K$ in Eq.~(\ref{Aint}) can be replaced by
$z_0^2$ and $K_0$. Inserting the expansions into Eqs.~(\ref{GG1}) and
(\ref{GG2}), we obtain a set of four coupled equations,
\begin{eqnarray}
\frac{dz_0^2}{dl} & = & 2z_0^2(2-\pi K_0) ,
\label{G1} \\
\frac{d\Delta z^2}{dl} & = & 2(2-\pi K_0) \Delta z^2 - 2\pi z_0^2
  \Delta K ,
\label{G2} \\
\frac{dK_0}{dl} & = & -4\pi^3 z_0^2 K_0^2 ,
\label{G3} \\
\frac{d\Delta K}{dl} & = & -4\pi^3 K_0^2 \Delta z^2 - 8\pi^3 z_0^2 K_0
  \Delta K - 4\pi^3 z_0^2 K_0^2 \nonumber \\
& & \times \exp\bigg[-8\pi^3 \int_0^l \! dl'\,
  z_0^2(l') K_0(l') \bigg] .
\label{Gp4}
\end{eqnarray}
These equations are subject to the boundary conditions
$z_0^2(0) = z^2(0) = \exp(2\beta\mu)$,
$\Delta z^2(0) = 0$,
$K_0(0) = K(0) = -\alpha_0(0)\beta q^2/2\pi$, and
$\Delta K(0) = 0$.
With the help of Eq.~(\ref{G3}) the integral (\ref{Aint}) is found to
be
\begin{equation}
{\cal A}^2(l) = {\cal A}_0^2 \left(\frac{K_0(l)}{K(0)}\right)^2 .
\label{Asimp}
\end{equation}
Thus we obtain a purely differential equation for $\Delta K$:
\begin{equation}
\frac{d\Delta K}{dl} = -4\pi^3 K_0^2 \Delta z^2 - 8\pi^3 z_0^2 K_0
  \Delta K - 4 \pi^3 z_0^2 \frac{K_0^4}{K^2(0)} .
\label{G4}
\end{equation}

Equations (\ref{G1}) to (\ref{G3}) and Eq.~(\ref{G4})
describe the layered system for small $s/4\lambda_{ab}$. Note that the
number of equations has been reduced from infinity to four. Note
further that Eqs.~(\ref{G1}) and (\ref{G3}) are
the original Kosterlitz recursion relations,\cite{Kost} which
describe the {\it uncoupled\/} layers and can be solved by themselves.

The solution of the Kosterlitz recursion relations is well known. For $T$
smaller than a critical temperature $T_c$ the fugacity $z_0$
converges exponentially to zero for large $l$, while the stiffness
goes to a finite value $K_0 \ge 2/\pi$. Thus the vortices are bound
in small vortex-antivortex pairs.
For $T > T_c$ the fugacity diverges
exponentially and $K_0$ goes to zero for $l\to\infty$.
Here, very many unbound vortices
exist. At $T_c$ the stiffness constant $K_0$
jumps from $2/\pi$ to zero; this is the famous ``universal jump''.

Let us now take the electromagnetic coupling between the layers into
account. The asymptotic behavior of the quantity ${\cal A}^2$ is
given by Eq.~(\ref{Asimp}), whereas the asymptotic forms of
$\Delta z^2$
and $\Delta K$ can be read off from Eqs.~(\ref{G2}) and (\ref{G4}) in
connection with the known forms of $z_0^2$ and $K_0$.

We first consider the case $T>T_c$. The quantity $\Delta z^2$ diverges
exponentially for
large $l$. Thus the fugacity $z=\sqrt{z_0^2+\Delta z^2 {\cal A}_0^2}$
also diverges exponentially, whereas $\Delta K$, $K$, and
the coupling parameter $\alpha_0$ vanish. Thus the fugacity and the
stiffness constant qualitatively behave as
in the case of a single layer. Note that
$d/dl\,(\Delta z^2/z_0^2)=-2\pi\Delta K$
so that $\Delta z^2/z_0^2$ approaches a constant for $l\to\infty$ and the
expansion (\ref{exp1}) remains valid for arbitrarily large $l$.
{}From Eq.~(\ref{Asimp})
it follows that ${\cal A}^2$ goes to zero in this regime. Since
${\cal A}^2$ is a measure of the electromagnetic coupling between the
layers this means that the renormalized interaction between vortices in
different layers vanishes faster than the interaction within the same
layer. In this sense the layers decouple and the system becomes
two-dimensional for $T>T_c$.

For $T<T_c$, on the other hand, the quantity $\Delta z^2$ and, thus,
the fugacity $z$ converge towards zero. The correction to the
stiffness, $\Delta K$, approaches a finite and negative value.
Therefore the coupling parameter $\alpha_0$ of
vortices within the same layer is reduced by the presence of other
layers, the correction being proportional to $s/\lambda_{ab}$.
As long as this correction is small, the pairs are still bound, though
less tightly. The quantitative significance of this correction is
discussed below. The quantity ${\cal A}^2$ is reduced by the
renormalization,
but remains finite. Thus the interaction between vortices in different
layers remains finite, and the system is three-dimensional for $T<T_c$.

Whereas the above results are obtained by analytical study of the
asymptotic behavior of the recursion relations, numerical
integration is necessary to produce trajectories in the
$\Delta K$-$\Delta z^2$ plane. Trajectories for $z(0)=0.03$ and
several starting values $K(0)$ are shown in Fig.~\ref{fig:DD}. The
value $z(0)=\exp(\beta\mu)=0.03$ seems reasonable for typical HTSC's
(Ref.~\onlinecite{Frie} gives the value $\mu=-E_c=-0.72 q^2$ and
$\beta q^2$ must be larger than 4).

We now turn to the question of the absolute size of the correction
$\Delta K$ to the stiffness constant below the critical temperature.
Numerical studies of Eqs.~(\ref{G2}) and (\ref{G4}) show that
\begin{equation}
\Delta K(\infty) \cong \frac{2\pi^3 z^2(0) K_0^2(\infty)}
  {2-\pi K_0(\infty)}
\label{eq103}
\end{equation}
is a very good approximation for small $z(0)$ and small
$2-\pi K_0(\infty)$.
The temperature dependence of $K_0$ is known to be\cite{Halp}
\begin{equation}
K_0(\infty) \cong \frac{2}{\pi} \left(1+\sqrt{2B}\sqrt{\frac{T_c-T}
  {T_c}}\right) ,
\end{equation}
where $B$ is a nonuniversal constant. Thus
\begin{equation}
\Delta K(\infty) \cong -2\pi z^2(0) \sqrt{\frac{2T_c}{B}} \:
  (T_c-T)^{-1/2}
\label{DKinf}
\end{equation}
to leading order in $T_c-T$.

If the critical temperature is approached from below, the correction to
the stiffness constant thus diverges with an exponent of $-1/2$. Very
near to the transition temperature of a single layer, $T_c$,
this divergence causes the stiffness
constant $K(\infty)$ to become negative, which is physically impossible.
In this region higher order terms neglected here should render
$K\ge 0$. We expect that $K$ vanishes and the pairs start to break up
at a temperature $T_c^{\text{3D}}<T_c$. $T_c^{\text{3D}}$ should be
larger than the temperature $T^\ast$ at which $K$ vanishes in the
linear approximation (which thus definitely fails at $T^\ast$). $T^\ast$
can be estimated from Eq.~(\ref{DKinf}).
For Bi-2212 the parameters are $z(0) \approx 0.03$,
$s/4\lambda_{ab} \approx 0.0019$, and
$\sqrt{2B} \approx 2.91$.\cite{Mart} We thus obtain
$T^\ast/T_c = 1-1.3\cdot 10^{-10}$.
Obviously, $T^\ast$ is indistinguishable from $T_c$, the linear
approximation is always valid in practice, and the transition
temperature $T_c^{\text{3D}}$ is not significantly shifted by the
electromagnetic coupling.

The foregoing discussion indicates that the effect of the other layers
is very small. Indeed, even if we could measure a correction of
$\Delta K/K_0 = 1\%$ in Bi-2212,
we would still need to resolve a temperature range
of the order of $10^{-6}\,T_c$ to see it, which is impossible.
In practice the effect of the electromagnetic coupling will be overruled
by the Josephson coupling, see below.

By using HTSC superlattices it may be possible to increase
the ratio $s/\lambda_{ab}$. An additional advantage of superlattices
is that the electromagnetic approximation is more appropriate in this
case.
However, even then the effect will be very small. More promising
systems are superlattices fabricated from {\it conventional\/}
superconductors.

\section{Layered superconductors with Josephson coupling}
\label{sec:Joe}

\subsection{General consequences of Josephson coupling}

Since Josephson tunneling between the layers has been found
experimentally\cite{Muel} in Bi-2212 and other HTSC's
we are faced with the question how to incorporate the Josephson
coupling into the renormalization group theory presented above.

The Josephson coupling between the layers leads to
the appearance of JV's. Because of the smoothness of the
phase of the
order parameter in the layers and the conservation of magnetic flux
every two-dimensional vortex (i.e.\ pancake vortex) in a layer
must be connected to two JV strings and vice versa.
Since the external magnetic field vanishes and surface effects are
neglected the JV's form {\it vortex loops\/}. On these loops pancake
vortices sit, where ever the loops penetrate a layer.
It is clear from this picture that a JV connects either
a vortex and an antivortex within the same layer or two vortices of
the same vorticity in neighboring layers.

The program of the renormalization group approach to this
particular system is to
derive an expression for the energy of any possible configuration
of JV's, add this expression to the Hamiltonian,
insert the new Hamiltonian into the partition function,
integrate out small vortex-antivortex pairs, and derive a set of
recursion relations. To obtain a tractable expression
for the energy of JV's we introduce several approximations.

The energy of JV strings is approximately proportional to their
length $L$ for $L\gtrsim\lambda_J$,\cite{Feig,Bu46,Timm} where
$\lambda_J$ is the Josephson length. If two pancakes are
separated by less than $\lambda_J$, no full JV develops and
the energy due to the Josephson coupling is not simply
linear.\cite{Frie}
However, for small separations the electromagnetic contribution to the
interaction dominates and the form for the Josephson term
is unimportant. Furthermore, as the interaction is renormalized, small
pancake pairs are integrated out and only large pairs remain, for
which the linear approximation is correct. Thus we assume that
the energy of a JV of length $L$ is given by
\begin{equation}
U_J = \kappa L .
\label{UJ}
\end{equation}
It is further assumed that the JV's form straight lines between the
pancakes. Thus we may take the JV's into account by adding
a term of the form (\ref{UJ}) to the Hamiltonian for any
two vortices that are connected by a JV.
The Hamiltonian takes the form
\begin{equation}
{\cal H} = \frac{1}{2} \sum_{i\not=j} q_iq_j \alpha_{n_i-n_j}
  \ln \frac{|{\bf r}_i-{\bf r}_j|}{\tau}
  + \sum_{(p,p')\in {\cal C}_J} \kappa\, |{\bf r}_{p'}-{\bf r}_p| ,
\label{HmJ}
\end{equation}
where ${\cal C}_J$ denotes a given configuration of JV's
connecting the pancakes $i=1,\ldots,2N$ and $(p,p')$ is a single
JV string characterized by the pancakes $p$ and $p'$ at its ends.

The constant $\kappa$ in Eq.~(\ref{HmJ}) can be evaluated for small
$s/\lambda_{ab}$, the unrenormalized value being\cite{Bu46}
$\kappa(l=0) \cong q^2 \lambda_{ab}/s\lambda_c$,
where $\lambda_c$ is the penetration depth for magnetic fields
applied parallel to the layers. Note that
$\lambda_{ab}/s\lambda_c \approx 10^{-3}\text{ \AA}^{-1}$ for Bi-2212.
We can now see that the electromagnetic approximation
is not sufficient for Bi-2212: The average distance between neighboring
pairs and the separation on which the linear term in the interaction
becomes important are of the same order of magnitude (1000 \AA{}).

\widetext
Since the partition function is a sum over all possible configurations,
we must, in addition to the summations and integrations in
Eq.~(\ref{Z0}), sum over all configurations of JV's for given pancake
positions. Thus the grand canonical partition function may be written
as
\begin{equation}
{\cal Z} = \sum_N \frac{1}{N!^2} \left(\frac{z}{\tau^2}\right)^{2N}
  \sum_{n_1} \int_{D_1} \!\!d^2r_1 \cdots \sum_{n_{2N}} \int_{D_{2N}}
  \!\!d^2r_{2N} \sum_{{\cal C}_J} \exp(-\beta {\cal H})
\label{ZJ}
\end{equation}
with ${\cal H}$ given by Eq.~(\ref{HmJ}). The last sum in
Eq.~(\ref{ZJ}) is over all configurations ${\cal C}_J$ of JV's.

We are now faced with the task to rewrite the prescription
(\ref{intout}), which tells us how to integrate out small pairs,
for the partition function (\ref{ZJ}).
The new aspect here is
that we must take all possibilities into account to insert the small
pair into the JV loops. These possible insertions are of three
topologically
different types: 1. the small pair can form a vortex loop by itself,
as shown in Fig.~\ref{fig:case123}(a). 2. the pair can be part of
a larger vortex loop, but still be directly connected by one
JV, see Fig.~\ref{fig:case123}(b). 3. the pair is not
directly connected by a JV, see Fig.~\ref{fig:case123}(c).
Since the present approach neglects some contributions to the interlayer
vortex correlations as discussed in subsection~\ref{suse:ren},
we may expect unphysical
results if vortex loops across more than one layer are essential.
On the other hand, the approach
is consistent if only small loops of type 1 are important.

If the sum over all possible insertions of the small pair $(i,j)$ into
the configuration ${\cal C}'_J$ of JV's between the other
pancakes is denoted by
\begin{equation}
\sum_{(i,j)\searrow {\cal C}'_J} ,
\label{inssum}
\end{equation}
the prescription for integrating out small pairs is given by
\begin{eqnarray}
\lefteqn{
\sum_{n_1} \int_{D_1} \!\!d^2r_1 \cdots \sum_{n_{2N}} \int_{D_{2N}}
  \!\!d^2r_{2N} \sum_{{\cal C}_J}
  \cong \sum_{n_1} \int_{D'_1} \!\!d^2r_1 \cdots
  \sum_{n_{2N}} \int_{D'_{2N}} \!\!d^2r_{2N} \sum_{{\cal C}_J} }
  \nonumber \\
& & \quad{}+\frac{1}{2} \sum_{i\not=j} \sum_{n_1} \int_{D'_1}
  \!\!d^2r_1 \cdots
  \sum_{n_{i-1}} \int_{D'_{i-1}} \!\!d^2r_{i-1} \sum_{n_{i+1}}
  \int_{D'_{i+1}}
  \!\!d^2r_{i+1} \cdots \sum_{n_{j-1}} \int_{D'_{j-1}} \!\!d^2r_{j-1}
  \sum_{n_{j+1}} \int_{D'_{j+1}} \!\!d^2r_{j+1} \cdots \nonumber \\
& & \qquad\times \sum_{n_{2N}} \int_{D'_{2N}} \!\!d^2r_{2N}
  \sum_{{\cal C}'_J}
  \sum_{n_j} \int_{\overline{D}_j} \!\!d^2r_j
  \sum_{n_i} \int_{\tau\le |{\bf r}_i-{\bf r}_j|<\tau+d\tau} \!\!\!\!
  d^2r_i \!\! \sum_{(i,j)\searrow {\cal C}'_J} \!\!
  \delta_{n_i,n_j} \delta_{q_i,-q_j} .
\label{ioJ}
\end{eqnarray}
Here, ${\cal C}'_J$ denotes a configuration of JV's if the
pancakes $i$ and $j$ are absent. Note that the sum (\ref{inssum})
contains one summand for the first type of vortex loops and
many summands for the second and the third type.
Before the general case is discussed, we turn to a simple limiting case.

\subsection{Very strong Josephson coupling}

Let us now consider a layered superconductor with {\it very strong\/}
Josephson
coupling between the layers. In this case the energy of the JV strings
is generally large compared to the other energy scales.
Fluctuations of the JV's cost much energy and
the system prefers states with small total JV length for any given
configuration of pancake vortices.
Although states with perfectly aligned pancakes in all layers have
the lowest Josephson energy, such states are not created by thermal
fluctuations since their total energy is infinite. The elementary
excitations are small vortex loops across one layer (see
Fig.~\ref{fig:case123}(a)).
Larger, more complicated loops will not form, except within a narrow
3D region, since the energy of JV strings is large.
Thus we consider only loops of the first type,
and the sum (\ref{inssum}) is reduced to one term.

The limit of large $\kappa$ is not physically accessible since
for a layered superconductors one always finds
$\lambda_c > \lambda_{ab}$.
Nevertheless it is considered here because it forms the opposite
limiting case as compared with the system without any Josephson
coupling and should allow one to estimate the maximum effect Josephson
coupling can have.

The additional energy resulting from the two JV's is
$2U_J = 2\kappa\,|{\bf r}_i-{\bf r}_j| = 2\kappa\tau$.
This is a {\it constant\/} term in the Hamiltonian.
Therefore, a constant factor $e^{-2\beta\kappa\tau}$
appears in the integrand.

For the same reasons as in the exclusively electromagnetic case only
pairs within the same layer can be integrated out. The calculations
are similar.
The constant factor mentioned above
is just dragged through. After the integration and the replacement
of $\tau$ by $\tau+d\tau$, we obtain the partition function
\begin{eqnarray}
{\cal Z} & = & \exp\!\left[2\pi\left(\frac{z}{\tau^2}\right)^2
  \tau\,d\tau MF e^{-2\beta\kappa\tau}\right]
  \sum_N \frac{1}{N!^2} \left(\frac{z}{(\tau+d\tau)^2}\right)^{2N}
  \nonumber \\
& & \times \left[1+\Bigl(2+\frac{\beta}{2}q^2\alpha_0\Bigr)
  \frac{d\tau}{\tau} \right]^{2N}
  \sum_{n_1} \int_{D'_1} \!\!d^2r_1 \cdots \sum_{n_{2N}} \int_{D'_{2N}}
  \!\!d^2r_{2N} \sum_{{\cal C}_J} \nonumber \\
& & \times \exp\Bigg[-\frac{\beta}{2} \sum_{i\not=j}
  \left(\alpha_{n_i-n_j}
  + 2\pi^2 z^2\frac{d\tau}{\tau} \beta q^2 \tilde{\alpha}_{n_i-n_j}
  e^{-2\beta\kappa\tau}\right) q_iq_j \ln\frac{|{\bf r}_i-{\bf r}_j|}
  {\tau+d\tau}
  - 2\beta\kappa \sum_{(p,p')\in {\cal C}_J}
  |{\bf r}_{p'}-{\bf r}_p| \Bigg] .
\label{ZEJ}
\end{eqnarray}
Again, there is no rescaling of lengths in the $z$ direction,
in contrast to Refs.\ \onlinecite{Pier} and
\onlinecite{Frie}. However, the arguments given in
subsection~\ref{suse:ren}
remain valid in the presence of Josephson coupling and, therefore, the
system is still essentially two-dimensional.

\narrowtext
Comparison with the original partition function (\ref{ZJ}) yields the
recursion relations
\begin{eqnarray}
\frac{dz^2}{dl} & = & z^2 (4+\beta q^2 \alpha_0) ,
\label{Gz2J} \\
\frac{d\alpha_n}{dl} & = & 2\pi^2 z^2 \beta q^2
  \exp(-2\beta\kappa\tau_0 e^l) \sum_m \alpha_{m-n} \alpha_m ,
\label{GanJ} \\
\frac{d\kappa}{dl} & = & 0 ,
\label{GkJ}
\end{eqnarray}
where again $l = \ln \tau/\tau_0$.
We thus find that the linear coupling constant $\kappa$ is not
renormalized due to screening if we consider only loops of the first type.
(Neither is there renormalization due to rescaling of lengths, as noted
above.) These equations,
especially Eq.\ (\ref{GkJ}), will break down very near to the phase
transition due to the appearance of larger, three-dimensional
vortex loops.

The additional factor $\exp(-2\beta\kappa\tau_0 e^l)$ in the equation
for $\alpha_n$ approaches zero very rapidly. The change of the
electromagnetic coupling parameters $\alpha_n$ with the scale $l$
and, therefore, the screening of the electromagnetic
interaction are strongly suppressed. In connection with the observation
that the linear interaction due to JV's is not screened at all within
this approximation, this result indicates that the low-temperature
phase of bound pairs is
stabilized and $T_c$ is higher than in the electromagnetic case.

To address this issue the recursion relations (\ref{Gz2J})
and (\ref{GanJ}) are expanded for small $s/4\lambda_{ab}$. The
procedure is analogous to the one employed above and
is not repeated here. It is seen that the {\it electromagnetic\/}
coupling between the layers brings about a very small correction
to the renormalized electromagnetic coupling within the same layer.
Thus it suffices to consider the vortex fugacity $z_0$ and the
stiffness constant $K_0$ to order zero. The renormalization of these
quantities is described by the equations
\begin{eqnarray}
\frac{dz_0^2}{dl} & = & 2z_0^2(2-\pi K_0) ,
\label{G1J} \\
\frac{dK_0}{dl} & = & -4\pi^3 z_0^2 K_0^2
  \exp(-2\beta\kappa\tau_0e^l) .
\label{G3J}
\end{eqnarray}
Note that $\beta q^2=2\pi K(l=0)$.
Numerical integration for $z(0)=0.03$,
$\kappa = 10^{-3}\text{ \AA}^{-1} q^2$,
$\tau_0=21.5\text{ \AA}$, and several values of $K(0)$ yields
the trajectories shown in Fig.~\ref{fig:BKTJ}.

By studying the asymptotic behavior of Eqs.~(\ref{G1J}) and (\ref{G3J})
two temperature regimes can be identified. In this section the
temperature which
divides these two regimes is denoted by $T_c$, whereas the original
BKT transition temperature is $T_{\text{BKT}}$. As is shown
in Fig.~\ref{fig:BKTJ}, $T_c$ is significantly higher than
$T_{\text{BKT}}$.
For Bi-2212, experiments on ultra-thin films have found a transition
temperature\cite{Roge}
$T_{\text{BKT}} = 35\text{ K}$ and a mean-field temperature\cite{Mart}
$T_{c0} = 86,8\text{ K}$. From the above results, the transition
temperature for bulk Bi-2212 should be $T_c = 37.4\text{ K}$, but
the experimental value is $84.7 \text{ K}$.\cite{Mart} Taking into
account that a more realistic description of the
Josephson coupling (inclusion of larger loops)
would probably yield an even lower $T_c$, the main
effect must be brought about by other mechanisms, e.g.\ change of the
hole density in the CuO$_2$ planes.

For $T<T_c$ the fugacity goes to zero and the stiffness constant
approaches a finite value $K_0(\infty) > 2/\pi$ as
\mbox{$l\to\infty$}. The linear coupling
due to JV's is not changed. The vortices are bound
in small vortex-antivortex pairs.

For $T>T_c$ the fugacity diverges exponentially, but the stiffness
constant
$K_0$ still approaches a non-vanishing value $K_0(\infty) < 2/\pi$.
Very many vortices exist, but they are still bound in pairs.
This result is different from the exclusively electromagnetic case.
Also, the electromagnetic interaction between vortices in
different layers does not vanish and the system remains
three-dimensional even above $T_c$.

Furthermore, numerical studies indicate that there is no special
feature
in $K_0(\infty)$ at $T_c$ (see Fig.~\ref{fig:BKTJ}). Nor do we expect the
appearance of ohmic resistance at $T_c$ since the vortices are bound in
pairs below as well as above this temperature. Thus the question arises
of whether there
really is a phase transition at $T_c$. However, since the true transition
is governed by effects not included here (larger, 3D loops), this
result should be interpreted with caution.

\subsection{Intermediate Josephson coupling---failure of the
renormalization group approach}

Let us finally consider a layered superconductor with general
Josephson coupling. In this case we must take all the
possible configurations of JV's into account
(see Figs.~\ref{fig:case123}). This is especially necessary in the
vicinity of $T_c$. Here, we consider only one term that
shows in what way the approach fails.

It suffices to investigate one summand of the sum (\ref{inssum})
of the second type (see Fig.~\ref{fig:case123}(b)).
Let ${\bf r}_0^i$ (${\bf r}_1^j$)
be the position of the neighbor of vortex $i$ ($j$) other than
vortex $j$ ($i$), see Fig.~\ref{fig:vorbez}.
The energy of these three JV's is given by
$\kappa\, |{\bf r}_i-{\bf r}_0^i| + \kappa\tau
+ \kappa\,|{\bf r}_j-{\bf r}_1^j|$. Furthermore, to insert the new pair
we must remove the JV from ${\bf r}_0^i$ to ${\bf r}_1^j$. Thus
we have to consider an additional factor
\begin{equation}
\exp\left(-\beta\kappa\,|{\bf r}_i-{\bf r}_0^i|-\beta\kappa\tau
-\beta\kappa\,|{\bf r}_j-{\bf r}_1^j|
+\beta\kappa\,|{\bf r}_1^j-{\bf r}_0^i|\right)
\end{equation}
in the integrand.
The integral over the lowest order term in $\beta$ of the factor
containing the electromagnetic interaction---this term is simply
unity---times the
above factor
is evaluated in the appendix. After integration over ${\bf r}_i$
and ${\bf r}_j$ we obtain the expression
\begin{eqnarray}
2\pi\tau\,d\tau\, \exp\left(-\beta\kappa\tau
  +\beta\kappa\,|{\bf r}_1^j-{\bf r}_0^i|\right)\,
  I_0(\beta\kappa\tau) \nonumber \\
\times |{\bf r}_1^j-{\bf r}_0^i|^2\,
  K_2(\beta\kappa|{\bf r}_1^j-{\bf r}_0^i|) ,
\end{eqnarray}
where $I_0$ and $K_2$ are Bessel functions. Expansion for large
separations $r=|{\bf r}_1^j-{\bf r}_0^i|$ yields
\begin{equation}
2\pi\tau\,d\tau\, e^{-\beta\kappa\tau}\, I_0(\beta\kappa\tau)
  \sqrt{\frac{\pi}{2\beta\kappa}}\: r^{3/2} .
\end{equation}
After performing the sums over the layer indices $n_i$
and $n_j$ and the vortex indices $i$ and $j$ we end up with
a partition function based on a Hamiltonian that contains a term
proportional to $|{\bf r}_{p'}-{\bf r}_p|^{3/2}$ for any two
pancakes $p$ und $p'$ that are connected by a JV.

It can be shown that this term cannot be canceled by any other terms:
All other terms in the integrand either contain factors $\beta q^2$ or
$\beta^2 q^4$ (as opposed to the term discussed above) or have the
same sign.

But no term proportional to the vortex separation to the power $3/2$ is
present in the original Hamiltonian! Thus it is impossible to
renormalize the
parameters in the partition function in such a way as to regain the
original partition function (\ref{ZJ}). The program of the
renormalization group approach thus fails for the layered
superconductor with (general) Josephson coupling, at least within the
present model. We speculate that the reason for this failure lies in
the one-dimensional nature of the JV strings as opposed to the
point-like pancake vortices.

\section{Conclusions}

The application of the renormalization group formulation of the
BKT theory\cite{Kost} to layered superconductors has been investigated
in detail. This program has successfully been carried out for a layered
system without Josephson coupling. It has been found that
the transition temperature $T_c$ is not affected by the presence of
other layers and that the
correction to the stiffness constant $K$ is significant only in a
inaccessibly narrow temperature range below $T_c$
for typical HTSC's, but may be observable for layered structures
made of conventional superconductors. The opposite limiting case
of very strong Josephson coupling has also been investigated. A
significant upward shift of the transition temperature has been found,
but even using approximations valid only
for unphysically strong Josephson coupling the shift is still
too small to account for the difference in the transition temperatures
between Bi-2212 bulk und thin film samples.

Finally we have shown that the renormalization group approach fails for
intermediate Josephson coupling. The reason for this failure has been
argued to lie in the one-dimensional nature of the Josephson vortex
strings. One would welcome a renormalization
group theory for this system, but if and how it
may be constructed remains a question for the future.

\acknowledgements

The author would like to thank Prof.\ J. Appel, A. Zabel, and
T. Wolenski for valuable discussions. Financial support by the
Deutsche Forschungsgemeinschaft is acknowledged.

\widetext

\appendix

\section{}
\label{app:B}

As discussed in Sec.~\ref{sec:Joe} we have to integrate the expression
$\exp(-\beta\kappa\,|{\bf r}_i-{\bf r}_0^i|
-\beta\kappa\,|{\bf r}_1^j-{\bf r}_j|)$ over ${\bf r}_i$ and
${\bf r}_j$. The integral over ${\bf r}_i$ is
given by
\begin{eqnarray}
{\cal I}_1 & = & \int_{\tau\le |{\bf r}_i-{\bf r}_j|<\tau+d\tau}
  \!\!\!\! d^2r_i \, \exp\left(-\beta\kappa\,|{\bf r}_i-{\bf r}_0^i|
  -\beta\kappa\tau -\beta\kappa\,|{\bf r}_1^j-{\bf r}_j|
  +\beta\kappa\,|{\bf r}_1^j-{\bf r}_0^i|
  \right) \nonumber \\
& = &  \tau\,d\tau
  \exp\left(-\beta\kappa\tau-\beta\kappa\,|{\bf r}_j-{\bf r}_1^j|
  +\beta\kappa\,|{\bf r}_1^j-{\bf r}_0^i|
  \right) \nonumber \\
& & \times \int_0^{2\pi} \!d\theta \exp\!\left(-\beta\kappa\,
  |{\bf r}_j-{\bf r}_0^i| \sqrt{1+2\frac{\tau}{|{\bf r}_j-{\bf r}_0^i|}
  \cos\theta + \frac{\tau^2} {|{\bf r}_j-{\bf r}_0^i|^2}}\, \right) .
\end{eqnarray}
Expansion for small $\tau/|{\bf r}_j-{\bf r}_0^i|$ yields
\begin{eqnarray}
{\cal I}_1 & \cong & \tau\,d\tau \exp\left(
  -\beta\kappa\tau-\beta\kappa\,|{\bf r}_j-{\bf r}_1^j|
  +\beta\kappa\,|{\bf r}_1^j-{\bf r}_0^i| \right)
  \int_0^{2\pi} \!d\theta\,\exp\!\left[-\beta\kappa
  \,|{\bf r}_j-{\bf r}_0^i|
  \left(1+\frac{\tau}{|{\bf r}_j-{\bf r}_0^i|}\cos\theta\right) \right]
  \nonumber \\
& = & 2\pi \tau\,d\tau \exp\left(-\beta\kappa
  \,|{\bf r}_j-{\bf r}_0^i|-\beta\kappa\tau
  -\beta\kappa\,|{\bf r}_j-{\bf r}_1^j|
  +\beta\kappa\,|{\bf r}_1^j-{\bf r}_0^i|
  \right)\, I_0(\beta\kappa) ,
\end{eqnarray}
and integration over ${\bf r}_j$ finally gives
\begin{equation}
\int_{\overline{D}_j} \!\! d^2r_j \,{\cal I}_1
  \cong  2\pi\tau\,d\tau\: \exp\left(-\beta\kappa\tau
  +\beta\kappa\,|{\bf r}_1^j-{\bf r}_0^i|\right)\,
  I_0(\beta\kappa)\:
  |{\bf r}_1^j-{\bf r}_0^i|^2 \:K_2\!\left(\beta\kappa\,
  |{\bf r}_1^j-{\bf r}_0^i|\right) .
\end{equation}
This is the result stated above.

\narrowtext


\begin{figure}
\caption{The trajectories of Eqs.~(\protect\ref{G2}) and
(\protect\ref{G4}), obtained by numerical integration for $z(0)=0.03$
and several values of $K(0)$. The flow is to the left.}
\label{fig:DD}
\end{figure}

\begin{figure}
\caption{The three distinct configuration of a small vortex-antivortex
pair with respect to the JV's. The arrows denote
pancake vortices and the thick gray lines JV's. See text.}
\label{fig:case123}
\end{figure}

\begin{figure}
\caption{The trajectories
for a system with strong Josephson coupling, but vanishing
electromagnetic coupling between the layers. The flow is to the
left. The dashed curve would be
the critical trajectory in the original BKT case.}
\label{fig:BKTJ}
\end{figure}

\begin{figure}
\caption{Definition of vortex coordinates. The circles denote pancake
vortices and the dashed line is a JV. The arrows point in the direction
of the magnetic field within the JV.}
\label{fig:vorbez}
\end{figure}

\end{document}